# Accurate Electron Affinities and Orbital Energies of Anions from a Non-Empirically Tuned Range-Separated Density Functional Theory Approach

*Lindsey N. Anderson, M. Belén Oviedo, and Bryan M. Wong\**

Department of Chemical & Environmental Engineering and Materials Science & Engineering Program, University of California-Riverside, Riverside, California 92521, United States

\*Corresponding author. E-mail: bryan.wong@ucr.edu. Homepage: http://www.bmwong-group.com

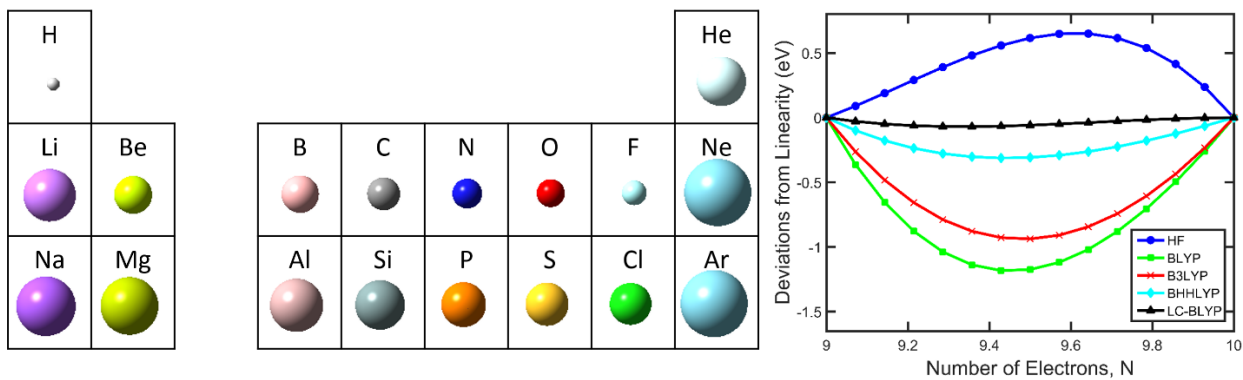

**TOC graphic**

ABSTRACT. The treatment of atomic anions with Kohn-Sham density functional theory (DFT) has long been controversial since the highest occupied molecular orbital (HOMO) energy, $E_{HOMO}$, is often calculated to be positive with most approximate density functionals. We assess the accuracy of orbital energies and electron affinities for all three rows of elements in the periodic table (H-Ar) using a variety of theoretical approaches and customized basis sets. Among all of the theoretical methods studied here, we find that a non-empirically tuned range-separated approach (constructed to satisfy DFT-Koopmans' theorem for the *anionic* electron system) provides the best accuracy for a variety of basis sets – even for small basis sets where most functionals typically fail. Previous approaches to solve this conundrum of positive $E_{HOMO}$ values have utilized non-self-consistent methods; however electronic properties, such as electronic couplings/gradients (which require a self-consistent potential and energy), become ill-defined with these approaches. In contrast, the non-empirically tuned range-separated procedure used here yields well-defined electronic couplings/gradients and correct $E_{HOMO}$ values since both the potential and resulting electronic energy are computed self-consistently. Orbital energies and electron affinities are further analyzed in the context of the electronic energy as a function of electronic number (including fractional numbers of electrons) to provide a stringent assessment of self-interaction errors for these complex anion systems.

**Introduction**

The quantum mechanical description of weakly-bound anions and their unusual properties continues to garner immense interest in the atomic/molecular physics and condensed-phase chemistry communities. In particular, the weak binding of an extra electron to a stable neutral atom/molecule is central to the study of Rydberg states,[1-3] few-body quantum systems,[4] and their couplings to the electronic continuum.[5-7] Within the rapidly-growing field of condensed-phase chemistry, loosely-bound electrons are present as solvated electrons in which an extra electron is not associated to any one particular molecule but is collectively bound by a cluster of solvent molecules.[8-10] In the broader fields of chemistry and materials science, anions and radicals play a vital role in semiconducting molecular clusters,[11,12] fullerenes,[13,14] charge transfer,[15] and solar cells.[16,17]

To describe these highly complex processes and electronic environments, an accurate quantum mechanical treatment is necessary, and advances in density functional theory (DFT) have enabled first-principles calculations with reasonable accuracy (mean absolute errors around 0.2 eV).[18] However, during the formative stages of DFT in quantum chemistry, serious concerns were raised that Kohn-Sham DFT was not appropriate for the study of anions since the highest occupied molecular orbital (HOMO) energy was often calculated to be positive – typically for small basis sets, and surprisingly for atomic species with sizeable electron affinities.[19-23] A positive HOMO eigenvalue is problematic since this implies that the anion is unbound, and the calculation is, in principle, unreliable.[24] At the time, these problematic cases were shown to arise from the deficiency of LDA/GGA exchange-correlation functionals since Kohn-Sham potentials obtained from these approximations exhibit the wrong asymptotic behavior.[25] Still, other researchers have argued that electron affinities calculated using reasonable basis sets (even with these conventional

DFT methods) are still reliable when calculated as a difference between the self-consistent total energies of the anion and neutral species.[26] These controversial issues continue to interest DFT purists, and recent parallel studies by Burke[27] and Jensen[28] have demonstrated that anions which should be bound states in reality are actually described as metastable electronic resonances by many approximate DFT methods. Burke and co-workers[27] have attributed these discrepancies to strong self-interaction errors (SIEs) that arise from a net negative charge, resulting in an effective potential where the last electron is actually unbound.

Numerous publications have appeared using DFT methods to calculate electron affinities[18, 29-31] (including recent developments in extended Koopmans' Theorem approaches[32-34]), and three solutions have been proposed to address the conundrum of positive $E_{HOMO}$ values in Kohn-Sham DFT: (1) *Ignore warnings on positive HOMO energies:*[18] This viewpoint is a pragmatic approach to simply "march on" and compute electron affinities as energy differences between the anion and neutral species; however, the presence of a formally problematic, positive HOMO energy is unsettling since one "obtains the right answer for the wrong reason"; (2) *Use an orbital-dependent self-interaction correction scheme:*[35, 36] This approach can, in principle, directly eliminate SIE terms in an orbital-by-orbital procedure; however, a detailed study by Scuseria and co-workers[23] has shown that these self-interaction corrections can severely impair equilibrium properties, in addition to introducing computational difficulties due to the invariance of the energy with respect to unitary transformations; (3) *Self-consistently compute orbitals using Hartree-Fock (HF), but use the HF density to non-self-consistently evaluate the energy with DFT:*[27, 37] While this most recently proposed approach will formally result in a bound anion with a negative $E_{HOMO}$, the calculation of electronic properties, such as electronic couplings or gradients, becomes ill-defined since the potential is evaluated with one approach while the energy is non-self-consistently evaluated with another approximation.

In this work, we instead propose the following alternative: use a non-empirically tuned procedure to satisfy DFT-Koopmans' theorem for the anionic ($N$+1) electron system, and use the resulting tuned XC functional to *self-consistently* evaluate *both* orbital energies and electron affinities. This procedure should yield correct $E_{HOMO}$ values and well-defined electronic couplings/gradients since both the potential and resulting electronic energy are computed self-consistently. We test the accuracy of this approach by computing both the orbital energies and electron affinities for all three rows of elements in the periodic table (H-Ar). We have chosen to focus our attention on individual atoms since SIEs are particularly severe for isolated atoms where the extra electron is strongly localized.[38] A variety of theoretical methods and extremely diffuse, customized basis sets are used in this work (containing exponents less than $10^{-10}$), which require special modifications to existing codes to achieve convergence at the basis set limit. Finally, we examine the orbital energies and electron affinities in the context of the electronic energy, $E$, as a function of electronic number, $N$, including fractional numbers of electrons.[39] These tests of deviations from linearity provide a stringent assessment of SIEs inherent to the underlying functional itself as well as a critical diagnostic of the basis set used in the calculation.[40] We give a detailed analysis of these $E$ vs. $N$ curves and discuss the implications of using a non-empirically tuned approach for obtaining accurate and formally-correct bound anions with well-defined electronic properties in a fully self-consistent approach.

**Theory and Methodology**

The main purpose of this work is to (1) assess the accuracy of non-empirically tuned range-separated DFT and (2) understand the effects of non-standard, *extremely* diffuse basis sets for simultaneously computing electron affinities and orbital energies for anions – both of which are briefly reviewed here.

**Non-Empirically Tuned Range-Separated DFT.** In contrast to conventional hybrid functionals, the range-separated formalism[24, 25] mixes short range density functional exchange with long range Hartree-Fock exchange by partitioning the electron repulsion operator into short and long range terms (i.e., the mixing parameter is a function of electron coordinates):

$$\frac{1}{r_{12}} = \frac{1-\text{erf}(\mu \cdot r_{12})}{r_{12}} + \frac{\text{erf}(\mu \cdot r_{12})}{r_{12}}. \tag{1}$$

The erf term denotes the standard error function, $r_{12}$ is the interelectronic distance between electrons 1 and 2, and $\mu$ is the range-separation parameter in units of Bohr$^{-1}$. The second term in Eq. (1) is of particular importance since it enforces a rigorously correct 100% contribution of asymptotic HF exchange, which we[41-45] and others[46, 47] have found to be essential for accurately describing long-range charge-transfer excitations, orbital energies, and valence excitations in even relatively simple molecular systems. For pure density functionals (such as the generalized gradient BLYP used here or the PBE kernel) which do not already include a fraction of nonlocal Hartree-Fock exchange, the exchange-correlation (XC) energy within the range-separated formalism is

$$E_{xc}(\mu) = E_{c,\text{DFT}}(\mu) + E_{x,\text{DFT}}^{\text{SR}}(\mu) + E_{x,\text{HF}}^{\text{LR}}(\mu). \tag{2}$$

$E_{c,\text{DFT}}$ is the DFT correlation functional, $E_{x,\text{DFT}}^{\text{SR}}$ is the short-range DFT exchange functional, and $E_{x,\text{HF}}^{\text{LR}}$ is the Hartree-Fock contribution to exchange computed with the long-range part of the Coulomb operator. Baer and Kronik[48, 49] have shown that the range-separation parameter, $\mu$, is system dependent but can be non-empirically tuned to satisfy DFT-Koopmans' theorem.[50-52] In summary, this theorem states that the energy of the highest occupied molecular orbital (HOMO) equals the negative of the ionization potential (IP); the latter is typically obtained from the difference of two separate energy calculations via a ΔSCF procedure. Within the Kohn-Sham DFT formalism, this condition is fulfilled for the exact XC-functional; therefore, adjusting the range-separation parameter in this self-consistent manner provides a theoretical justification for this procedure. While the original non-empirical tuning procedure focused on computing the HOMO and IP for neutral systems, we have slightly modified this procedure to *non-empirically* compute electron affinities and orbital energies of anions by minimizing the following objective function:

$$J^2(\mu) = \left[\varepsilon_{\text{HOMO}}^{\mu}(N+1) + \text{EA}^{\mu}(N)\right]^2, \qquad (3)$$

where $\varepsilon_{\text{HOMO}}^{\mu}(N+1)$ is the HOMO energy of the *anionic* $(N+1)$-electron system and $\text{EA}^{\mu}(N)$ is the electron affinity computed via a ΔSCF energy difference between the $N$ and $N+1$ electron systems with the same range-separation parameter: $\text{EA}^{\mu}(N) = E^{\mu}(N) - E^{\mu}(N+1)$. Note that the electron affinity, $\text{EA}^{\mu}(N)$, used in this work is defined as a positive number, which is the same convention used by Jensen,[28] Burke,[27] and Tozer[40] in their previous studies of electron affinities. It should also be noted that with this definition, the electron affinity of the $N$-electron system is equal to the ionization potential of the $(N+1)$-electron system (i.e. $\text{EA}^{\mu}(N) = \text{IP}^{\mu}(N+1)$), so the objective function in Eq. (3) is a non-empirical approach to satisfy DFT-Koopmans' theorem for anions. To obtain the optimal $\mu$ values for each oligomer, several single-point energy calculations for each of the individual anions were carried out by varying $\mu$ from 0.1 to 0.6 (in increments of 0.05) for each of the $N$ and $N-1$ electron states. The objective function, $J^2$ (Eq. 3), was computed as a function of $\mu$ for each anion, and spline interpolation was subsequently used to refine the minimum for each individual system.

**Non-Standard, Extremely Diffuse Basis Sets.** Throughout this work, we compare two basis sets for computing the electron affinities and orbital energies in atomic anions: a conventional aug-pc-2 basis[53] (a triple-zeta quality basis augmented with diffuse functions) and a customized diffuse basis set which we denote as aug-pc-∞. Following the same approach as Jensen,[28] this customized aug-pc-∞ basis set was constructed by adding both diffuse $s$- and $p$- functions to the aug-pc-4 basis by scaling the outer exponent in a geometric progression with a factor of $\sqrt{10}$ until the exponent of the most diffuse function became less than $10^{-10}$. Furthermore, we have used the aug-pc-2 and aug-pc-∞ basis sets in their uncontracted forms to avoid any possible contraction errors. The uncontracted basis sets for all the anions are listed in the Supporting Information for completeness. As demonstrated by Jensen,[28] calculations with extremely diffuse basis functions of this magnitude pose a variety of numerical issues that standard settings in many computational chemistry codes are not capable of handling. As such, all calculations were carried out with a locally modified version of the Gaussian 09 package[54] specifically tailored to handle these non-standard, extremely diffuse basis sets. Specifically, all integral thresholds were tightened to machine precision and density matrices were converged to at least $10^{-8}$. The XC potential was calculated using an extremely dense Euler-Maclaurin radial grid[55] with 5000 points in combination with a Lebedev angular grid with 434 points by setting int(acc2e=20,grid=5000434) in the Gaussian route section. Previous work by Jensen[28] has shown that these large radial and angular grids are necessary for numerical integration involving basis functions having small exponents such that the Davidson radial norm criterion is fulfilled to within $10^{-6}$. In addition, threshold screenings for discarding integration points with low density were disabled by changing the constant `SmlExp` from 1.0d-6 to 1.0d-15 in the Gaussian routines `utilnz.F` and `l301.F`. Most importantly, we[56] and others[28] have found that XC-functionals containing a large percentage of exact exchange can converge to saddle points in the electronic parameter space, especially when near-degenerate orbitals are present[56] or when extremely diffuse basis sets are used.[28] As such, all SCF solutions were verified to be genuine minima in the electronic parameter space by carrying out a stability analysis to converge (if possible) toward a lower-energy broken-symmetry solution (by setting both scf(qc,conver=9) and stable=opt in the Gaussian route section), which allows for an unrestricted spin state as well as a reduction in symmetry of the orbitals.

**Results and Discussion**

Fig. 1 shows the smooth curves that result from computing $J^2$ as a function of $\mu$ for the first three rows of elements in the periodic table (H – Ar). The upper panel (a) in Fig. 1 depicts the $J^2$ plots using the smaller aug-pc-2 basis, and the lower panel (b) shows the corresponding plots for the same atoms with the customized diffuse aug-pc-∞ basis. The optimally-tuned $\mu$ values for all anions obtained with both basis sets are summarized in Table 1.

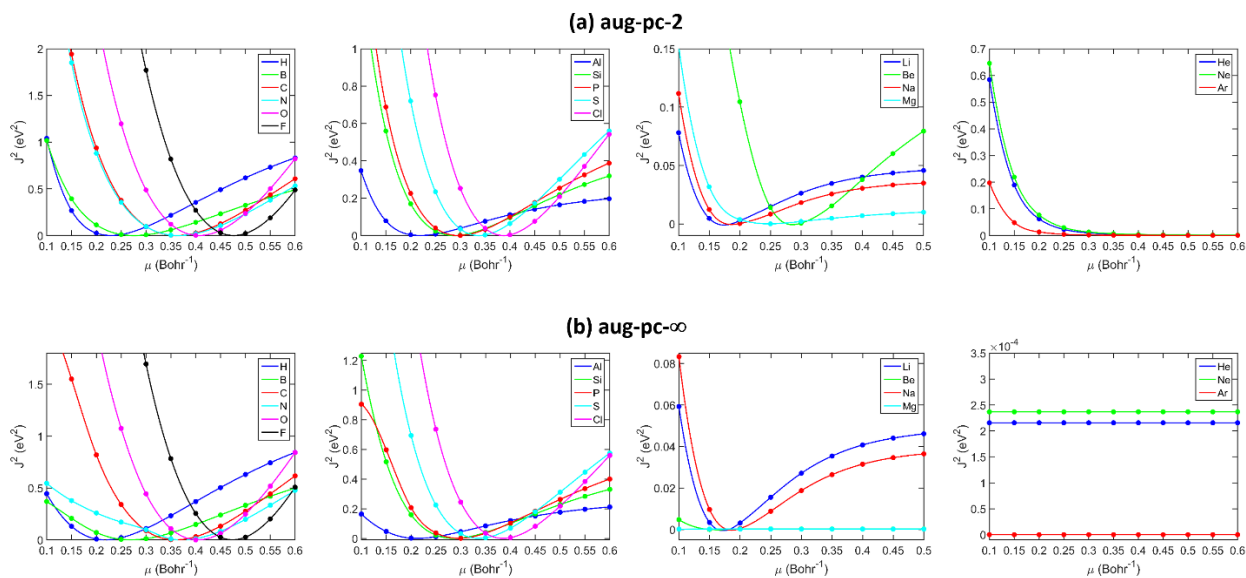

**Figure 1.** Plots of $J^2$ as a function of $\mu$ for all three rows of elements in the periodic table (H-Ar) using (a) the smaller aug-pc-2 and (b) the customized diffuse aug-pc-∞ basis sets.

| Anion | aug-pc-2 | aug-pc-∞ |
|---|---|---|
| H$^-$ | 0.22939 | 0.21759 |
| He$^-$ | – | – |
| Li$^-$ | 0.17396 | 0.17209 |
| Be$^-$ | 0.28572 | – |
| B$^-$ | 0.27586 | 0.26903 |
| C$^-$ | 0.36178 | 0.35958 |
| N$^-$ | 0.36519 | 0.37228 |
| O$^-$ | 0.40676 | 0.40425 |
| F$^-$ | 0.47631 | 0.47418 |
| Ne$^-$ | – | – |
| Na$^-$ | 0.18633 | 0.18614 |
| Mg$^-$ | 0.24821 | – |
| Al$^-$ | 0.21966 | 0.21218 |
| Si$^-$ | 0.28778 | 0.28643 |
| P$^-$ | 0.29623 | 0.29661 |
| S$^-$ | 0.33761 | 0.33593 |
| Cl$^-$ | 0.38736 | 0.38556 |
| Ar$^-$ | – | – |

**Table 1.** Non-empirically-tuned $\mu$ values for all three rows of elements in the periodic table (H-Ar) using (a) the aug-pc-2 and (b) the customized diffuse aug-pc-∞ basis set.

Based on the tabulated values of $\mu$ and their corresponding plots in Fig. 1, it is interesting to note that increasing the diffuseness of the basis set has a negligible effect (less than 1%) on the non-empirically tuned range-separated parameter. Minor exceptions to this trend include the Be and Mg atoms which exhibit $J^2$ minima for the smaller aug-pc-2 basis but show nearly flat lines for the customized diffuse aug-pc-∞ basis. This discrepancy arises since the Be and Mg atoms have a closed shell electronic configuration, and the addition of an extra electron leads to a strong inter-electronic repulsion. As such, the customized diffuse aug-pc-∞ basis possesses enough flexibility to allow the extra electron to drift off to infinity (discussed further at the end of this section), whereas the extra electron in the smaller aug-pc-2 basis is artificially confined to remain close to the nucleus. It should also be noted that both the aug-pc-2 and aug-pc-∞ basis do not exhibit $J^2$ minima for any of the noble gas atoms (He, Ne, and Ar) as both basis sets correctly predict these anions to be unbound. To further understand the sensitivity of our results to the range-separation parameter, $\mu$, we carried out a series of benchmark tests, presented in full detail in the Supporting Information. In short, we also investigated the more commonly-used IP-tuning procedure by minimizing the following objective function: $J^2(\mu) = \left[\varepsilon_{HOMO}^{\mu}(N) + IP^{\mu}(N)\right]^2$, where $\varepsilon_{HOMO}^{\mu}(N)$ is the HOMO energy of the neutral $N$-electron system and $IP^{\mu}(N)$ is the ionization potential computed via a $\Delta$SCF energy difference between the $N$ - $1$ and $N$ electron systems. With this common tuning procedure, Fig. SI-1 in the Supporting Information shows that the resulting $J^2$ plots are qualitatively different than the curves shown in Fig. 1 within the main text. In particular, the optimal $\mu$ values obtained with the IP-tuning procedure are larger ($\mu \sim 0.5$) than the corresponding $\mu$ values obtained with the EA-tuning procedure in Eq. (3). As a result, these benchmark results support our rationale for using the objective function in Eq. (3) for accurately calculating electron affinities, as opposed to the more commonly-used IP-tuning procedure typically used for neutral systems.

Finally, before we compare these results to other XC functionals, it is worth noting that the short-range DFT exchange in Eq. (2) decays exponentially on a length scale of $\sim 1/\mu$ and, therefore, smaller non-empirically tuned $\mu$ values are associated with larger systems (i.e., a smaller value of $\mu$ enables the short-range Coulomb operator to fully decay to zero on the length scale of the system). To demonstrate this trend, we plot the spatial extent, $\langle R^2 \rangle$ (i.e., the expectation value of $R^2$), for each anion as a function of $1/\mu$ obtained with the extremely diffuse aug-pc-$\infty$ basis (the $\langle R^2 \rangle$ vs. $1/\mu$ plot obtained with the smaller aug-pc-2 basis is similar and is given in the Supporting Information). Indeed, the optimal $\mu$ values generally reflect these trends and follow a nearly linear behavior (with the exception of the small H⁻ anion) with the larger-sized anions having smaller values of $\mu$ than smaller anions.

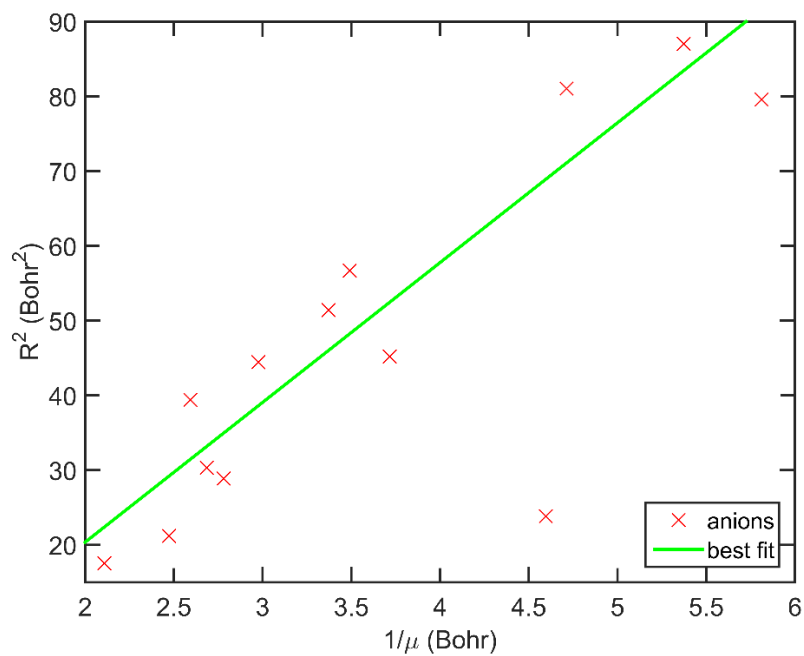

**Figure 2.** Plot of the spatial extent, $\langle R^2 \rangle$ (expectation value of $R^2$), for each anion as a function of $1/\mu$ obtained with the customized diffuse aug-pc-∞ basis. The outlier value near $1/\mu = 0.46$ (which corresponds to the H⁻ anion) was excluded when creating the best fit line.

Turning our attention to previous reports of unbound anions and anomalous orbital energies in conventional XC functionals, we compare our non-empirically tuned results to electron affinities and HOMO energies obtained with the BLYP, B3LYP, BHHLYP, and HF approaches. We have chosen these specific methods for comparison due to their widespread use in previous findings[27,28] and also because they represent different extremes of global hybrids where the HF exchange contribution ranges from 0.20 to 1.00. For example, the pure BLYP functional contains no HF exchange, the popular B3LYP hybrid functional is parameterized with 20% exchange, the half-and-half BHHLYP functional is constructed with 50% exchange, and the HF method is defined with 100% exchange. Moreover, a comparison of these XC functionals allows a fair and consistent evaluation since all of the DFT methods have similar LYP correlation[57] portions. It is worth noting that we also obtained similar results with other range-separated functionals such as LC-BOP, LC-

PBE, and LC-ωPBE, which is consistent with previous work by us[42-44] and Jensen[28] indicating that the long-range $E_{x,\text{HF}}^{\text{LR}}$ exchange term in Eq. (2) plays a more dominant role compared to either $E_{c,\text{DFT}}$ or $E_{x,\text{DFT}}^{\text{SR}}$ in these systems. We do not consider the heavily-parameterized Minnesota functionals in our study since the emphasis of this work is on *non-empirically* tuned functionals, and very recent reports have shown that most of the Minnesota functionals do not obey rigorously known constraints of the exact functional.[58, 59] Table 2 summarizes the electron affinities computed via a ΔSCF procedure (using the same EA = $E(N) - E(N+1)$ expression defined in the Theory and Methodology section) for the first three rows of elements in the periodic table. Total energies calculated with BLYP, B3LYP, BHHLYP, HF, and LC-BLYP for all three rows of elements in the periodic table are given in the Supporting Information. We emphasize that these non-standard calculations were incredibly difficult to converge, and the same integral thresholds, grid sizes, and computational settings described in the Theory and Methodology section were used. In addition, all SCF solutions were verified to be genuine minima in the electronic parameter space by carrying out a stability analysis to converge toward a lower-energy broken-symmetry solution.[56] Electron affinities for all of the noble gas atoms (He, Ne, and Ar) computed with the smaller aug-pc-2 basis were not included since these basis sets are unable to describe these unbound anions. Fig. 3 presents a graphical analysis of Table 2 by plotting the corresponding error (EA − EA$_{\text{expt.}}$) in the electron affinity.

**Table 2.** Electron affinities (in eV) computed via a ΔSCF procedure, EA = E(N) – E(N + 1), for all three rows of elements in the periodic table using the aug-pc-2 and customized diffuse aug-pc-∞ basis sets.

| Anion | EA$_{expt}$[a] | HF aug-pc-2 | HF aug-pc-∞ | BLYP aug-pc-2 | BLYP aug-pc-∞ | B3LYP aug-pc-2 | B3LYP aug-pc-∞ | BHHLYP aug-pc-2 | BHHLYP aug-pc-∞ | LC-BLYP aug-pc-2 | LC-BLYP aug-pc-∞ |
|---|---|---|---|---|---|---|---|---|---|---|---|
| H$^-$ | 0.75 | -0.28 | -0.33 | 0.85 | 1.21 | 0.91 | 1.09 | 0.67 | 0.71 | 0.83 | 0.84 |
| He$^-$ | 0.00 | – | 0.00 | – | 0.08 | – | 0.07 | – | 0.02 | – | 0.01 |
| Li$^-$ | 0.62 | -0.06 | -0.03 | 0.45 | 0.65 | 0.55 | 0.65 | 0.43 | 0.46 | 0.50 | 0.50 |
| Be$^-$ | 0.00 | -0.97 | -0.01 | -0.62 | 0.38 | -0.53 | 0.29 | -0.69 | 0.07 | -0.61 | 0.02 |
| B$^-$ | 0.28 | -0.30 | -0.30 | 0.46 | 0.92 | 0.47 | 0.76 | 0.18 | 0.35 | 0.42 | 0.43 |
| C$^-$ | 1.26 | 0.46 | 0.45 | 1.37 | 1.72 | 1.38 | 1.53 | 1.04 | 1.03 | 1.39 | 1.39 |
| N$^-$ | 0.00 | -1.64 | 0.00 | 0.40 | 1.14 | 0.25 | 0.81 | -0.25 | 0.23 | 0.23 | 0.23 |
| O$^-$ | 1.46 | -0.55 | -0.57 | 1.85 | 2.22 | 1.70 | 1.85 | 1.13 | 1.11 | 1.83 | 1.83 |
| F$^-$ | 3.40 | 1.21 | 1.17 | 3.70 | 3.80 | 3.55 | 3.53 | 2.92 | 2.90 | 3.76 | 3.74 |
| Ne$^-$ | 0.00 | – | 0.00 | – | 0.11 | – | 0.09 | – | 0.02 | – | 0.02 |
| Na$^-$ | 0.55 | -0.05 | -0.05 | 0.49 | 0.69 | 0.58 | 0.67 | 0.45 | 0.47 | 0.53 | 0.53 |
| Mg$^-$ | 0.00 | -0.58 | 0.00 | -0.42 | 0.23 | -0.31 | 0.19 | -0.45 | 0.05 | -0.45 | 0.01 |
| Al$^-$ | 0.43 | 0.03 | 0.01 | 0.39 | 0.71 | 0.47 | 0.65 | 0.28 | 0.36 | 0.35 | 0.36 |
| Si$^-$ | 1.39 | 0.87 | 0.85 | 1.24 | 1.43 | 1.35 | 1.40 | 1.16 | 1.15 | 1.26 | 1.25 |
| P$^-$ | 0.75 | -0.32 | -0.32 | 0.91 | 1.24 | 0.97 | 1.13 | 0.72 | 0.74 | 0.90 | 0.90 |
| S$^-$ | 2.08 | 0.90 | 0.88 | 2.14 | 2.25 | 2.21 | 2.21 | 1.96 | 1.95 | 2.19 | 2.19 |
| Cl$^-$ | 3.61 | 2.39 | 2.37 | 3.58 | 3.58 | 3.68 | 3.67 | 3.44 | 3.42 | 3.68 | 3.67 |
| Ar$^-$ | 0.00 | – | 0.00 | – | 0.11 | – | 0.09 | – | 0.02 | – | 0.00 |
| MAE[b] | | 1.03 | 0.69 | 0.21 | 0.33 | 0.17 | 0.23 | 0.24 | 0.14 | 0.20 | 0.11 |

[a]Experimental electron affinities from References 60-62. [b]Mean Absolute Errors relative to experimental electron affinities.

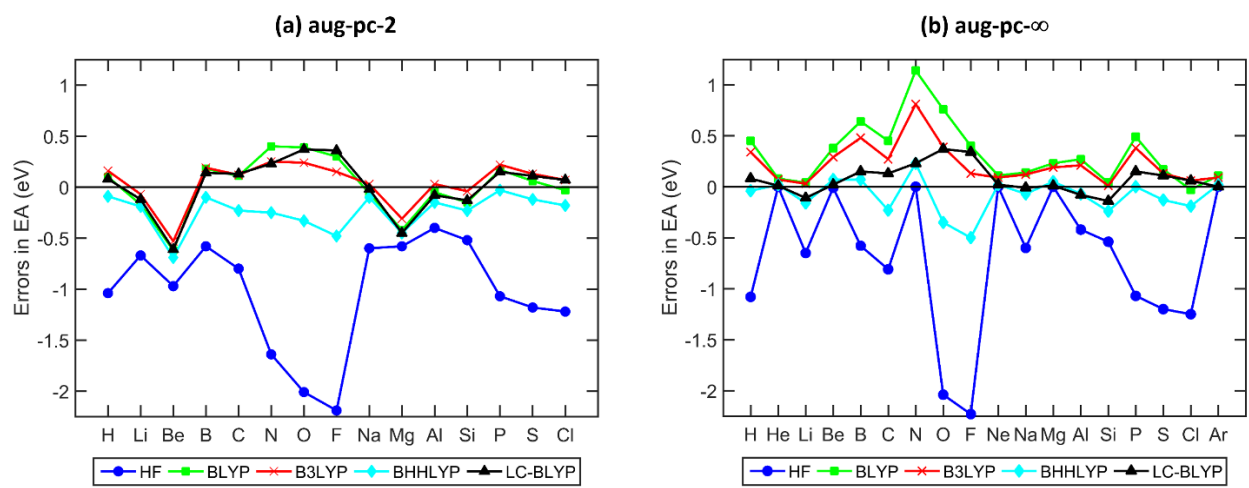

**Figure 3.** Errors in the electron affinity computed via a ΔSCF procedure, EA = E(N) – E(N + 1), for all three rows of elements in the periodic table using the (a) aug-pc-2 and (b) customized diffuse aug-pc-∞ basis sets.

Based on the tabulated mean absolute errors (MAE) for the customized aug-pc-∞ basis set, the non-empirically tuned LC-BLYP functional gives the best prediction of the ΔSCF electron affinities with a MAE of 0.11 eV, followed by the BHHLYP and B3LYP functionals with MAEs of 0.14 and 0.23 eV, respectively. The pure BLYP functional yields a larger MAE of 0.33 eV as this method does not contain any nonlocal exchange, whereas the HF method significantly underestimates electron affinities for several atoms (cf. Fig. 3), resulting in an even larger MAE of 0.69 eV due to its lack of explicit electron correlation. For all levels of theory and basis sets examined in this work, the largest errors in ΔSCF-computed electron affinities were observed in the halogen atoms (F and Cl) which arise from the strong electron repulsion that the extra electron feels when it is added to a nearly complete *p*-shell. In general, both Table 2 and Fig. 3 clearly demonstrate that XC functionals containing a larger percentage of HF exchange yield the most accurate electron affinities; however, there is an intricate balance between exchange and correlation, and incorporating only 100% nonlocal exchange without correlation as in the HF method (or conversely, neglecting nonlocal exchange completely as in the pure BLYP functional) can severely corrupt the prediction of electron affinities.

We now turn our attention to the central issue of unbound anions and positive HOMO energies originally mentioned in the Introduction. Table 3 summarizes the electron affinities computed from the *negative* HOMO energy of the anion, $-E_{\text{HOMO}}$ for the first three rows of elements in the periodic table. Again, electron affinities for all of the noble gas atoms (He, Ne, and Ar) computed with the smaller aug-pc-2 basis were not included since these basis sets are unable to describe these unbound anions. Fig. 4 presents a graphical analysis of Table 3 by plotting the corresponding error ($-E_{\text{HOMO}} - \text{EA}_{\text{expt.}}$) in the electron affinity. It is interesting to note that while the plots in Figs. 3 and 4 were calculated using two different metrics (ΔSCF vs. $-E_{\text{HOMO}}$), the resulting plots obtained with the LC-BLYP functional are nearly identical since the non-empirically tuned approach was constructed to satisfy the constraint in Eq. (3) as closely as possible. As mentioned in the Introduction, the *sign* of $E_{HOMO}$ for anions is a more stringent test of these XC functionals since a positive $E_{HOMO}$ implies that the anion is formally unbound. As clearly shown in Table 3, the non-empirically tuned LC-BLYP functional correctly predicts negative $E_{HOMO}$ values for *both* basis sets (note that Table 3 lists $-E_{HOMO}$ values) and for all anions, with the exception of the Be, Mg, and noble gas atoms – all of which *are* experimentally unbound. To highlight the robustness of these LC-BLYP results, we also carried out two benchmark tests in the Supporting Information to (1) ascertain the effect of using other basis sets and (2) compare electron affinities obtained with a *fixed* value of $\mu$ against the non-empirically tuned LC-BLYP results listed in Table 3. Tables SI-1 and SI-2 in the Supporting Information give the total energies and electron affinities calculated with the LC-BLYP functional (fixed at $\mu = 0.3$) for all three rows of elements in the periodic table using the aug-cc-pVTZ, aug-pc-2, and customized diffuse aug-pc-∞ basis sets. We have chosen $\mu = 0.3$ for our benchmark tests since 0.3 is the numerical average of the $\mu$ values listed in Table 1 for all three rows of elements in the periodic table (H-Ar). As demonstrated by the total energies in Table SI-1, Jensen's aug-pc-2 basis set is marginally better than Dunning's triple-zeta aug-cc-pVTZ basis, with the aug-pc-2 basis providing slightly lower total energies across the board. Interestingly, our benchmark test in Table SI-2 demonstrates that the electron affinities calculated with a single value of $\mu$ (= 0.3) exhibit impressively low MAE values that are only marginally worse (~ 0.04 eV larger in error) than the non-empirically tuned LC-BLYP results. Most importantly, the LC-BLYP results (fixed at $\mu = 0.3$) are *still* considerably more accurate than the electron affinities obtained from the other conventional functionals. To further demonstrate the applicability of this approach beyond single atoms, we have also calculated the electron affinities of molecules in the G2-1 data set, which was previously examined in a communication by Burke and co-workers.[37] Table SI-3 in the Supporting Information lists $-E_{HOMO}$ for molecular anions in the G2-1 set computed with the BLYP, B3LYP, BHHLYP, HF, and LC-BLYP approaches. In agreement with previous work by Burke and co-workers,[37] most molecular species (except $Cl_2$) are predicted to have positive $E_{HOMO}$ values with conventional DFT methods (i.e., BLYP, B3LYP, and BHHLYP), indicating an incorrect unbound nature with these functionals. Only HF and LC-BLYP correctly predict negative $E_{HOMO}$ values in Table SI-3; however, the HF values are significantly overestimated, resulting in a large MAE of 0.73 eV. In contrast, the LC-BLYP functional gives the lowest MAE of 0.29 eV and correctly predicts negative $E_{HOMO}$ values for all molecular species, which highlights the robustness of the range-separation approach in general. Returning to our original discussion on non-empirically tuned approaches for atoms, it should be noted that while the other functionals also give negative $E_{HOMO}$ values in Table 3 for the extremely diffuse aug-pc-∞ basis, their mean absolute errors are still significantly higher than the non-empirically tuned LC-BLYP (MAE = 0.12 eV) functional: HF has a MAE of 0.31 eV, followed by the BHHLYP and B3LYP functionals with MAEs of 0.62 and 0.87 eV. The pure BLYP functional exhibits the largest MAE of 0.92 eV when the aug-pc-∞ basis is used since almost all of the anions predicted by this functional are unbound or nearly unbound.

**Table 3.** Electron affinities (in eV) computed from the negative HOMO energy of the anion, $-E_{HOMO}$, for all three rows of elements in the periodic table using the aug-pc-2 and customized diffuse aug-pc-∞ basis sets.

| Anion | $EA_{expt}^{60-62}$ | HF aug-pc-2 | HF aug-pc-∞ | BLYP aug-pc-2 | BLYP aug-pc-∞ | B3LYP aug-pc-2 | B3LYP aug-pc-∞ | BHHLYP aug-pc-2 | BHHLYP aug-pc-∞ | LC-BLYP aug-pc-2 | LC-BLYP aug-pc-∞ |
|---|---|---|---|---|---|---|---|---|---|---|---|
| H⁻ | 0.75 | -0.07 | 1.26 | -1.76 | 0.02 | -1.00 | 0.01 | -0.11 | 0.01 | 0.83 | 0.84 |
| He⁻ | 0.00 | – | 0.00 | – | 0.02 | – | 0.01 | – | 0.01 | – | 0.03 |
| Li⁻ | 0.62 | 0.62 | 0.67 | -1.02 | 0.00 | -0.59 | 0.00 | -0.21 | 0.00 | 0.50 | 0.50 |
| Be⁻ | 0.00 | -0.49 | 0.00 | -2.48 | 0.01 | -2.36 | 0.00 | -1.73 | 0.00 | -0.62 | 0.03 |
| B⁻ | 0.28 | 0.77 | 0.78 | -1.83 | 0.03 | -1.28 | 0.02 | -0.62 | 0.01 | 0.42 | 0.43 |
| C⁻ | 1.26 | 2.12 | 2.13 | -1.98 | 0.01 | -1.10 | 0.01 | -0.01 | 0.01 | 1.39 | 1.39 |
| N⁻ | 0.00[a] | -1.46 | 0.00 | -2.95 | 0.00 | -2.22 | 0.00 | -1.39 | 0.00 | 0.23 | 0.23 |
| O⁻ | 1.46 | 2.19 | 2.17 | -2.35 | 0.00 | -1.26 | 0.00 | 0.19 | 0.21 | 1.83 | 1.83 |
| F⁻ | 3.40 | 4.93 | 4.92 | -1.43 | 0.01 | 0.02 | 0.05 | 1.97 | 1.97 | 3.76 | 3.74 |
| Ne⁻ | 0.00 | – | 0.00 | – | 0.03 | – | 0.02 | – | 0.01 | – | 0.03 |
| Na⁻ | 0.55 | 0.56 | 0.58 | -1.02 | 0.00 | -0.59 | 0.00 | -0.21 | 0.00 | 0.52 | 0.53 |
| Mg⁻ | 0.00 | -0.42 | 0.00 | -1.89 | 0.04 | -1.49 | 0.03 | -1.16 | 0.02 | -0.45 | 0.03 |
| Al⁻ | 0.43 | 0.60 | 0.61 | -1.40 | 0.01 | -0.93 | 0.00 | -0.45 | 0.00 | 0.35 | 0.36 |
| Si⁻ | 1.39 | 1.69 | 1.69 | -1.31 | 0.01 | -0.61 | 0.01 | 0.17 | 0.17 | 1.26 | 1.25 |
| P⁻ | 0.75 | 0.64 | 0.63 | -1.79 | 0.00 | -1.09 | 0.00 | -0.29 | 0.00 | 0.90 | 0.91 |
| S⁻ | 2.08 | 2.32 | 2.33 | -1.17 | 0.00 | -0.26 | 0.00 | 0.84 | 0.84 | 2.19 | 2.18 |
| Cl⁻ | 3.61 | 4.09 | 4.09 | -0.33 | 0.00 | 0.79 | 0.79 | 2.17 | 2.17 | 3.68 | 3.67 |
| Ar⁻ | 0.00 | – | 0.00 | – | 0.00 | – | 0.00 | – | 0.00 | – | 0.00 |
| MAE[b] | | 0.54 | 0.31 | 2.75 | 0.92 | 2.04 | 0.87 | 1.16 | 0.62 | 0.20 | 0.12 |

[a]Experimental electron affinities from References 60-62. [b]Mean Absolute Errors relative to experimental electron affinities.

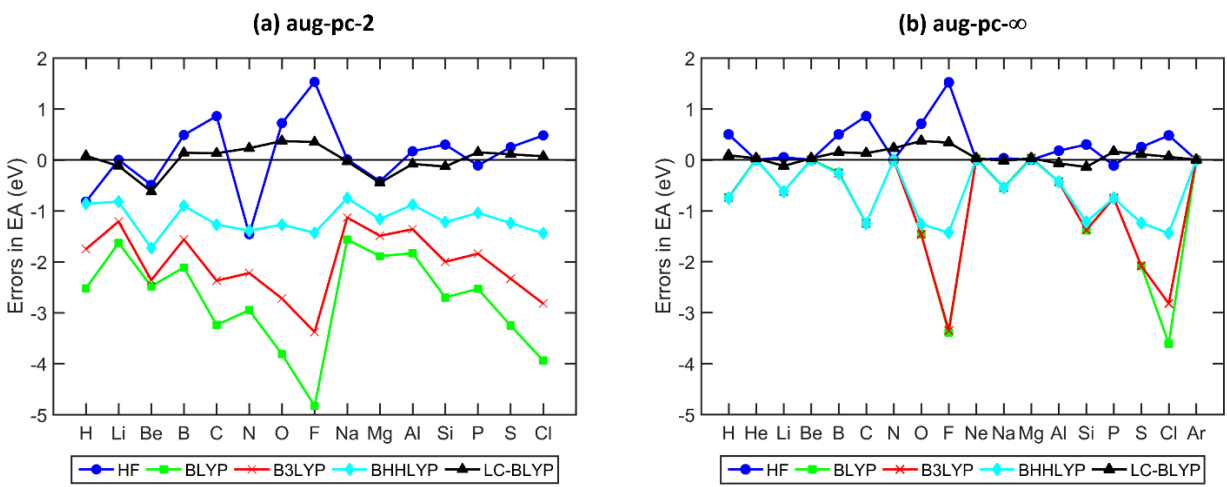

**Figure 4.** Errors in the electron affinity computed from the negative HOMO energy of the anion, $-E_{HOMO}$, for all three rows of elements in the periodic table using the (a) aug-pc-2 and (b) customized diffuse aug-pc-∞ basis basis sets.

In stark contrast to the qualitatively good trends obtained with the aug-pc-∞ basis, we encounter particularly worrisome results for the conventional XC functionals when the smaller

aug-pc-2 basis is used. While the non-empirically tuned LC-BLYP functional correctly gives negative $E_{HOMO}$ values for *both* basis sets, the *majority* of the aug-pc-2 $E_{HOMO}$ values predicted by *all* other DFT (not including HF) functionals are positive, incorrectly implying that these anions are formally unbound. These positive $E_{HOMO}$ values severely affect their accuracy, resulting in MAEs ranging from 1.16 eV (BHHLYP) to as high as 2.75 eV (BLYP) for the smaller aug-pc-2 basis set. While the purely nonlocal HF method does give negative $E_{HOMO}$ values, the MAEs obtained with this approach are nearly twice that of the non-empirically tuned LC-BLYP approach. Recognizing that the HF method correctly gives negative $E_{HOMO}$ values, Burke and co-workers[27] recently proposed the following solution to compute electron affinities: compute the orbitals using HF (where $E_{HOMO}$ is correctly predicted to be negative) but use the HF density to non-self-consistently evaluate the energy with an XC potential that contains both exchange and correlation. While this procedure will formally give a properly bound anion with a negative $E_{HOMO}$, various complications naturally arise: the calculation of electronic gradients becomes ill-defined since the electronic potential is evaluated with one approach while the energy is non-self-consistently evaluated with another approximation. Instead, we propose the following alternative: use a non-empirically tuned procedure (such as Eq. (3)) to satisfy DFT-Koopmans' theorem for the anionic ($N$+1) electron system, and use the resulting tuned XC functional to *self-consistently* evaluate *both* orbital energies and electron affinities. As a result, this procedure should yield correct $E_{HOMO}$ values and well-defined electronic gradients since both the potential and resulting electronic energy are computed self-consistently.

Finally, to complete our analysis of orbital energies and electron affinities of anions, we present a deeper analysis of the electronic energy as a function of electronic number (including fractional numbers of electrons). For an exact functional, Janak proved that $E$ is a piecewise linear function of $N$, with derivative discontinuities at integer number of electrons.[39] As such, a test of this deviation from linearity provides a stringent assessment of self-interaction errors inherent to the underlying functional itself as well as a diagnostic analysis of the basis set used in the calculation, as pointed out by a recent study by Tozer and co-workers[40] (discussed at the end of this section). Fig. 5 plots the variation of the electronic energy, $E$, as a function of electron number, $N$, for one piece of the $E(N)$ curve applied to the fluorine atom using the aug-pc-2 and customized aug-pc-∞ basis sets. Both the variation of $E$ vs. $N$, (a) and (c), as well as the deviation from linearity, (b) and (d), are also shown in Fig. 5. We have chosen to focus on fluorine since there have been strong discussions about this anion in several previous studies;[24, 26, 40] however, the main qualitative results discussed in this section equally apply to the other atomic anions.

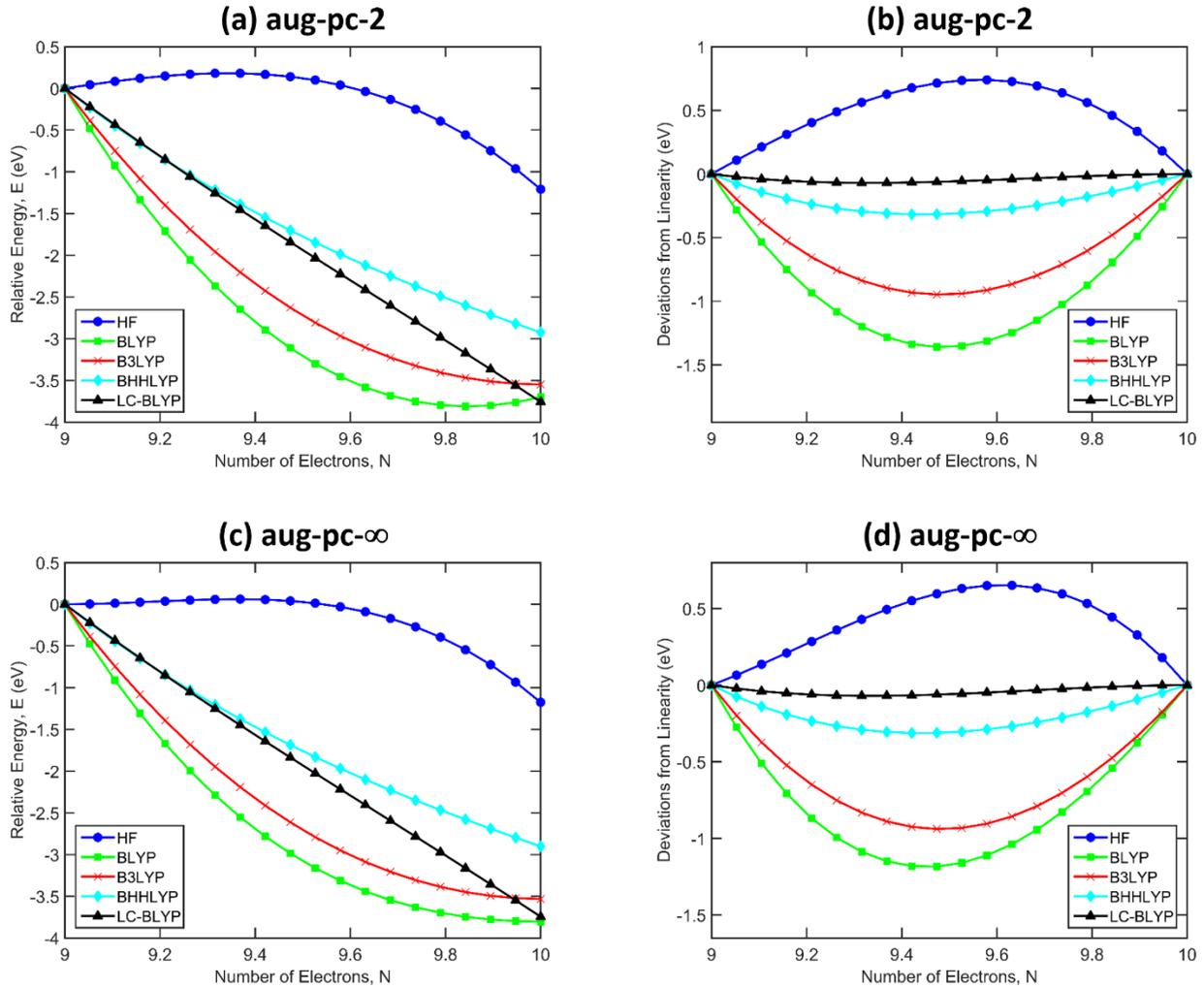

**Figure 5.** Variation of the electronic energy, $E$, as a function of electron number, $N$, applied to the fluorine anion using the (a)-(b) aug-pc-2 and (c)-(d) customized diffuse aug-pc-∞ basis sets. The variation of $E$ vs. $N$ is shown in (a) and (c), whereas the deviation from linearity is shown in (b) and (d). The nearly-exact straight line obtained with the non-empirically tuned LC-BLYP functional (for all plots (a)-(d)) implies that this method is nearly SIE-free for the fluorine anion.

As expected, the HF method yields a concave $E(N)$ curve and an energetic destabilization at fractional charges. In contrast, the pure BLYP and hybrid B3LYP/BHHLYP functionals tend to overdelocalize the anion, leading to a convex $E(N)$ curve and an over-stabilization (i.e. a lower energy) of the system at fractional charges. The non-empirically tuned LC-BLYP functional yields a nearly-exact straight line as a function of the fractional charge in the system (cf. Figs. 5(a) and

(c)). Furthermore, plots of the deviation from linearity in Figs. 5(b) and (d) imply that the non-empirically tuned LC-BLYP is nearly SIE-free for the fluorine anion.

As we conclude this discussion, it is worth mentioning a few subtle points concerning basis sets that can be seen in Figs. (a) and (c), which were recently discussed in a study by Tozer and co-workers.[40] For any functional, Janak proved that the slope of the $E(N)$ curve at integer number of electrons is related to the orbital energies. Specifically, the limiting value of $\partial E/\partial N$ to the left of an integer $M$ is the HOMO energy, $\varepsilon_{\text{HOMO}}(M)$:

$$\lim_{\delta \to 0} \left.\frac{\partial E}{\partial N}\right|_{N=M-\delta} = \varepsilon_{\text{HOMO}}(M), \tag{4}$$

and the limiting value of $\partial E/\partial N$ to the right of an integer $M$ is the LUMO energy, $\varepsilon_{\text{LUMO}}(M)$:

$$\lim_{\delta \to 0} \left.\frac{\partial E}{\partial N}\right|_{N=M+\delta} = \varepsilon_{\text{LUMO}}(M). \tag{5}$$

For the case of the fluorine system shown in Fig. 5, the $N = 9$ endpoint corresponds to the electron number of the neutral fluorine atom, and $N = 10$ is the electron number of the anion. Fig. 5(a) shows that the limiting value of $\partial E/\partial N$ to the left of $N = 10$ for the BLYP functional is slightly positive; consequently Eq. (4) states that the HOMO energy of the fluorine anion is also positive at the BLYP/aug-pc-2 level of theory, and the anion is formally unbound. However, the $E(N)$ curve for BLYP also possesses a shallow minimum at $N = 9.85$ in Fig. 5(a), and the energy of the fluorine anion could, in principle, be further stabilized by reducing the electron number to $N = 9.85$ with the remaining 0.15 fraction of an electron drifting off to infinity with zero energy. As pointed out by Tozer and co-workers,[40] the use of a small basis set prevents this from happening since all of the electrons are artificially confined to remain close to the nucleus. However, if the basis was sufficiently augmented with diffuse functions to enable fractional electron loss, the $E(N)$ curve would approach an idealized flattening and give $\varepsilon_{HOMO} = 0$ according to Eq. (4). Indeed, we do observe an idealized flattening of the *E*(*N*) curve for the BLYP functional in Fig. 5(c), leading to a nearly zero $\varepsilon_{HOMO}$ value (cf. Table 3) when the aug-pc-∞ basis is utilized. This same assessment, in regards to the flattening of the *E*(*N*) curve, can also be applied to the numerical BLYP results in Table 2, where the electron affinities computed via the ΔSCF procedure are observed to increase when the aug-pc-2 basis is systematically increased towards the larger aug-pc-∞ basis set. Using a similar analysis for concave behavior, the *E*(*N*) curve for HF possesses a shallow maximum at *N* = 9.35 in Fig. 5(a), and the system could, in principle, lower its energy by reducing the electron number in that vicinity to *N* = 9; however, a small basis set again prevents this. When the extremely diffuse aug-pc-∞ basis is used, the HF curve flattens near *N* = 9 and approaches the idealized behavior described by Tozer and co-workers. With this analysis, we now point out that the non-empirically tuned LC-BLYP functional does not exhibit any tendency towards a flattening or fractional electron loss (similar to the CAM-B3LYP analyses in Ref. 40), resulting in an accurate fluorine orbital energy/electron affinity *and* a correctly bound anion for *both* basis sets.

**Conclusions**

In this study, we have assessed the accuracy of orbital energies and electron affinities for all three rows of elements in the periodic table (H-Ar) using a variety of theoretical approaches and customized basis sets. Specifically, we have closely examined the electron affinities of these anions using two different metrics, ΔSCF vs. –$E_{HOMO}$, to understand the accuracy and intrinsic limitations of each theoretical approach. Among all of the theoretical methods studied here, we find that a non-empirically tuned range-separated approach (constructed to satisfy DFT-

Koopmans' theorem for the anionic ($N$ + 1)-electron system) provides the best accuracy for *both* metrics (ΔSCF and −$E_{HOMO}$) *as well as for both basis sets*. In contrast, the electron affinities obtained from "converged" $E_{HOMO}$ calculations with conventional XC functionals and smaller basis sets exhibit severe problems – the *majority* of these $E_{HOMO}$ values are predicted to be positive, incorrectly implying that these anions are formally unbound. While the purely nonlocal HF method does give negative $E_{HOMO}$ values and bound anions, there is a delicate balance between exchange and correlation, and the lack of electron correlation in HF yields errors that are nearly twice that of a non-empirically tuned range-separated approach (which correctly balances short range correlation with long-range exchange by satisfying DFT-Koopmans' theorem).

To address this conundrum of positive $E_{HOMO}$ values, other researchers have suggested to compute the orbitals using HF (where $E_{HOMO}$ is correctly predicted to be negative) but use the HF density to non-self-consistently evaluate the energy with an XC potential. While this procedure will formally result in a bound anion with a negative $E_{HOMO}$, the calculation of electronic properties, such as electronic couplings or gradients, becomes ill-defined since the potential is evaluated with one approach while the energy is non-self-consistently evaluated with another approximation. We instead propose the following alternative: use a non-empirically tuned procedure to satisfy DFT-Koopmans' theorem for the anionic ($N$+1) electron system, and use the resulting tuned XC functional to *self-consistently* evaluate *both* orbital energies and electron affinities. This procedure should yield correct $E_{HOMO}$ values and well-defined electronic couplings/gradients since the potential (and, hence, the electronic energy) is obtained from the derivative of an energy functional. Finally, we examine the orbital energies and electron affinities in the context of the electronic energy, $E$, as a function of electronic number, $N$, including fractional numbers of electrons. We find that the non-empirically tuned LC-BLYP functional yields a nearly-exact straight line as a function of the fractional charge in the system, and plots of the deviation from linearity imply that the non-empirically tuned LC-BLYP is nearly SIE-free. Moreover, a deeper analysis of the $E$ vs. $N$ curves demonstrates that the non-empirically tuned LC-BLYP functional does not exhibit any tendency towards a flattening or fractional electron loss, resulting in anions that are accurately described and correctly bound for *both* basis sets. Taken together, these calculations and analyses provide a natural methodology for obtaining accurate and formally-correct bound anions with well-defined electronic properties (such as electronic gradients and couplings) in a fully self-consistent approach.

**ASSOCIATED CONTENT**

**Supporting Information**

Plot of $<R^2>$ vs. $1/\mu$ obtained with the smaller aug-pc-2 basis, total energies for all three rows of elements (H – Ar) of the periodic table calculated at the HF, BLYP, B3LYP, BHHLYP, and LC-BLYP levels of theory with the aug-pc-2 and aug-pc-∞ basis sets, and uncontracted aug-pc-2 and aug-pc-∞ basis sets for all three rows of elements (H – Ar) of the periodic table.

**AUTHOR INFORMATION**

**Corresponding Author**

E-mail: *bryan.wong@ucr.edu.

**Notes**

The authors declare no competing financial interest.

**ACKNOWLEDGMENTS**

We gratefully acknowledge Prof. Frank Jensen for providing the computational details and settings required to converge SCF calculations with non-standard, extremely diffuse basis sets. This work was supported by the U.S. Department of Energy, Office of Science, Early Career Research Program under Award No. DE-SC0016269. We acknowledge the National Science Foundation for the use of supercomputing resources through the Extreme Science and Engineering Discovery Environment (XSEDE), Project No. TG- ENG160024.

**REFERENCES**

1. Watson, J. K. G. Effects of a Core Electric Dipole Moment on Rydberg States. *Mol. Phys.* **1994**, *81*, 277-289.
2. Kay, J. J.; Coy, S. L.; Petrović, V. S.; Wong, B. M.; Field, R. W. Separation of Long-Range and Short-Range Interactions in Rydberg States of Diatomic Molecules. *J. Chem. Phys.* **2008**, *128*, 194301.
3. Coy, S. L.; Grimes, D. D.; Zhou, Y.; Field, R. W.; Wong, B. M. Electric Potential Invariants and Ions-in-Molecules Effective Potentials for Molecular Rydberg States. *J. Chem. Phys.* **2016**, *145*, 234301.
4. Mitroy, J.; Safronova, M. S.; Charles, W. C. Theory and Applications of Atomic and Ionic Polarizabilities. *J. Phys. B: At., Mol. Opt. Phys.* **2010**, *43*, 202001.
5. Marante, C.; Argenti, L.; Martín, F. Hybrid Gaussian–*B*-Spline Basis for the Electronic Continuum: Photoionization of Atomic Hydrogen. *Phys. Rev. A* **2014**, *90*, 012506.
6. Lopata, K.; Govind, N. Near and Above Ionization Electronic Excitations with Non-Hermitian Real-Time Time-Dependent Density Functional Theory. *J Chem. Theory Comput.* **2013**, *9*, 4939-4946.
7. Sommerfeld, T.; Ehara, M. Complex Absorbing Potentials with Voronoi Isosurfaces Wrapping Perfectly Around Molecules. *J. Chem. Theory Comput.* **2015**, *11*, 4627-4633.
8. Verlet, J. R. R.; Bragg, A. E.; Kammrath, A.; Cheshnovsky, O.; Neumark, D. M. Observation of Large Water-Cluster Anions with Surface-Bound Excess Electrons. *Science* **2005**, *307*, 93-96.
9. Siefermann, K. R.; Liu, Y.; Lugovoy, E.; Link, O.; Faubel, M.; Buck, U.; Winter, B.; Abel, B. Binding Energies, Lifetimes and Implications of Bulk and Interface Solvated Electrons in Water. *Nat. Chem.* **2010**, *2*, 274-279.
10. Tang, Y.; Shen, H.; Sekiguchi, K.; Kurahashi, N.; Mizuno, T.; Suzuki, Y.-I.; Suzuki, T. Direct Measurement of Vertical Binding Energy of a Hydrated Electron. *Phys. Chem. Chem. Phys.* **2010**, *12*, 3653-3655.

11. Taylor, T. R.; Asmis, K. R.; Xu, C.; Neumark, D. M. Evolution of Electronic Structure as a Function of Size in Gallium Phosphide Semiconductor Clusters. *Chem. Phys. Lett.* **1998**, *297*, 133-140.
12. Bundhun, A.; Abdallah, H. H.; Ramasami, P.; Schaefer, H. F. Germylenes: Structures, Electron Affinities, and Singlet−Triplet Gaps of the Conventional $XGeCY_3$ (X = H, F, Cl, Br, and I; Y = F and Cl) Species and the Unexpected Cyclic $XGeCY_3$ (Y = Br and I) Systems. *J. Phys. Chem. A* **2010**, *114*, 13198-13212.
13. Chang, A. H. H.; Ermler, W. C.; Pitzer, R. M. Carbon Molecule (C60) and its Ions: Electronic Structure, Ionization Potentials, and Excitation Energies. *J. Phys. Chem.* **1991**, *95*, 9288-9291.
14. Dreuw, A.; Cederbaum, L. S. Multiply Charged Anions in the Gas Phase. *Chem. Rev.* **2002**, *102*, 181-200.
15. Han, H.; Zimmt, M. B. Solvent-Mediated Electron Transfer:  Correlation Between Coupling Magnitude and Solvent Vertical Electron Affinity. *J. Am. Chem. Soc.* **1998**, *120*, 8001-8002.
16. Kim, J. Y.; Lee, K.; Coates, N. E.; Moses, D.; Nguyen, T.-Q.; Dante, M.; Heeger, A. J. Efficient Tandem Polymer Solar Cells Fabricated by All-Solution Processing. *Science* **2007**, *317*, 222-225.
17. Shoaee, S.; Clarke, T. M.; Huang, C.; Barlow, S.; Marder, S. R.; Heeney, M.; McCulloch, I.; Durrant, J. R. Acceptor Energy Level Control of Charge Photogeneration in Organic Donor/Acceptor Blends. *J. Am. Chem. Soc.* **2010**, *132*, 12919-12926.
18. Rienstra-Kiracofe, J. C.; Tschumper, G. S.; Schaefer, H. F.; Nandi, S.; Ellison, G. B. Atomic and Molecular Electron Affinities:  Photoelectron Experiments and Theoretical Computations. *Chem. Rev.* **2002**, *102*, 231-282.
19. Shore, H. B.; Rose, J. H.; Zaremba, E. Failure of the Local Exchange Approximation in the Evaluation of the $H^-$ ground state. *Phys. Rev. B* **1977**, *15*, 2858-2861.
20. Schwarz, K. Instability of Stable Negative Ions in the Xα Method or Other Local Density Functional Schemes. *Chem. Phys. Lett.* **1978**, *57*, 605-607.
21. Sen, K. D. Instability of Stable Negative Ions in Xα Method. *Chem. Phys. Lett.* **1980**, *74*, 201-202.
22. Cole, L. A.; Perdew, J. P. Calculated Electron Affinities of the Elements. *Phys. Rev. A* **1982**, *25*, 1265-1271.
23. Vydrov, O. A.; Scuseria, G. E. Ionization Potentials and Electron Affinities in the Perdew–Zunger Self-Interaction Corrected Density-Functional Theory. *J. Chem. Phys.* **2005**, *122*, 184107.
24. Jarcki, A. A.; Davidson, E. R. Density Functional Theory Calculations for $F^−$. *Chem. Phys. Lett.* **1999**, *300*, 44-52.
25. van Leeuwen, R.; Baerends, E. J. Exchange-Correlation Potential with Correct Asymptotic Behavior. *Phys. Rev. A* **1994**, *49*, 2421-2431.
26. Galbraith, J. M.; Schaefer III, H. F. Concerning the Applicability of Density Functional Methods to Atomic and Molecular Negative Ions. *J. Chem. Phys.* **1996**, *105*, 862-864.
27. Lee, D.; Furche, F.; Burke, K. Accuracy of Electron Affinities of Atoms in Approximate Density Functional Theory. *J. Phys. Chem. Lett.* **2010**, *1*, 2124-2129.
28. Jensen, F. Describing Anions by Density Functional Theory: Fractional Electron Affinity. *J. Chem. Theory Comput.* **2010**, *6*, 2726-2735.

29. Chattaraj, P. K.; Duley, S. Electron Affinity, Electronegativity, and Electrophilicity of Atoms and Ions. *J. Chem. Eng. Data* **2010**, *55*, 1882-1886.
30. Gong, L.; Wu, X.; Li, W.; Qi, C.; Xiong, J.; Guo, W. Structures and Electron Affinities of $BrO_2F_2$ and $BrO_2F_3$. *Mol. Phys.* **2009**, *107*, 701-709.
31. Feng, X.; Li, Q.; Gu, J.; Cotton, F. A.; Xie, Y.; Schaefer, H. F. Perfluorinated Polycyclic Aromatic Hydrocarbons: Anthracene, Phenanthrene, Pyrene, Tetracene, Chrysene, and Triphenylene. *J. Phys. Chem. A* **2009**, *113*, 887-894.
32. Bozkaya, U. The Extended Koopmans' Theorem for Orbital-Optimized Methods: Accurate Computation of Ionization Potentials. *J. Chem. Phys.* **2013**, *139*, 154105.
33. Bozkaya, U. Accurate Electron Affinities from the Extended Koopmans' Theorem Based on Orbital-Optimized Methods. *J. Chem. Theory Comput.* **2014**, *10*, 2041-2048.
34. Yildiz, D.; Bozkaya, U. Assessment of the Extended Koopmans' Theorem for the Chemical Reactivity: Accurate Computations of Chemical Potentials, Chemical Hardnesses, and Electrophilicity Indices. *J. Comput. Chem.* **2016**, *37*, 345-353.
35. Perdew, J. P.; Zunger, A. Self-Interaction Correction to Density-Functional Approximations for Many-Electron Systems. *Phys. Rev. B* **1981**, *23*, 5048-5079.
36. Perdew, J. P. Size-Consistency, Self-Interaction Correction, and Derivative Discontinuity in Density Functional Theory. A*dv. Quantum Chem.* **1990**, *21*, 113-134.
37. Kim, M.-C.; Sim, E.; Burke, K. Communication: Avoiding Unbound Anions in Density Functional Calculations. *J. Chem. Phys.* **2011**, *134*, 171103.
38. Rösch, N.; Trickey, S. B. Comment on "Concerning the Applicability of Density Functional Methods to Atomic and Molecular Negative Ions" [*J. Chem. Phys.* **1996**, *105*, 862-864]. *J. Chem. Phys.* **1997**, *106*, 8940-8941.
39. Janak, J. F. Proof that $\partial E/\partial n = \varepsilon$ in Density-Functional Theory. *Phys. Rev. B* **1978**, *18*, 7165-7168.
40. Peach, M. J. G.; Teale, A. M.; Helgaker, T.; Tozer, D. J. Fractional Electron Loss in Approximate DFT and Hartree–Fock Theory. *J. Chem. Theory Comput.* **2015**, *11*, 5262-5268.
41. Foster, M. E.; Wong, B. M. Nonempirically Tuned Range-Separated DFT Accurately Predicts Both Fundamental and Excitation Gaps in DNA and RNA Nucleobases. *J. Chem. Theory Comput.* **2012**, *8*, 2682-2687.
42. Wong, B. M.; Piacenza, M.; Sala, F. D. Absorption and Fluorescence Properties of Oligothiophene Biomarkers from Long-Range-Corrected Time-Dependent Density Functional Theory. *Phys. Chem. Chem. Phys.* **2009**, *11*, 4498-4508.
43. Wong, B. M.; Cordaro, J. G. Coumarin Dyes for Dye-Sensitized Solar Cells: A Long-Range-Corrected Density Functional Study. *J. Chem. Phys.* **2008**, *129*.
44. Wong, B. M.; Hsieh, T. H. Optoelectronic and Excitonic Properties of Oligoacenes: Substantial Improvements from Range-Separated Time-Dependent Density Functional Theory. *J. Chem. Theory Comput.* **2010**, *6*, 3704-3712.
45. Raeber, A. E.; Wong, B. M. The Importance of Short- and Long-Range Exchange on Various Excited State Properties of DNA Monomers, Stacked Complexes, and Watson–Crick Pairs. *J. Chem. Theory Comput.* **2015**, *11*, 2199-2209.
46. Richard, R. M.; Herbert, J. M. Time-Dependent Density-Functional Description of the 1La State in Polycyclic Aromatic Hydrocarbons: Charge-Transfer Character in Disguise? *J. Chem. Theory Comput.* **2011**, *7*, 1296-1306.

47. Kuritz, N.; Stein, T.; Baer, R.; Kronik, L. Charge-Transfer-Like π→π* Excitations in Time-Dependent Density Functional Theory: A Conundrum and Its Solution. *J. Chem. Theory Comput.* **2011**, *7*, 2408-2415.
48. Stein, T.; Eisenberg, H.; Kronik, L.; Baer, R. Fundamental Gaps in Finite Systems from Eigenvalues of a Generalized Kohn-Sham Method. *Phys. Rev. Lett.* **2010**, *105*.
49. Stein, T.; Kronik, L.; Baer, R. Reliable Prediction of Charge Transfer Excitations in Molecular Complexes Using Time-Dependent Density Functional Theory. *J. Am. Chem. Soc.* **2009**, *131*, 2818-2820.
50. Stein, T.; Kronik, L.; Baer, R. Prediction of Charge-Transfer Excitations in Coumarin-Based Dyes Using a Range-Separated Functional Tuned from First Principles. *J. Chem. Phys.* **2009**, *131*, 244119.
51. Kronik, L.; Stein, T.; Refaely-Abramson, S.; Baer, R. Excitation Gaps of Finite-Sized Systems from Optimally Tuned Range-Separated Hybrid Functionals. *J. Chem. Theory Comput.* **2012**, *8*, 1515-1531.
52. Stein, T.; Kronik, L.; Baer, R. Reliable Prediction of Charge Transfer Excitations in Molecular Complexes Using Time-Dependent Density Functional Theory. *J. Am. Chem. Soc.* **2009**, *131*, 2818-2820.
53. Jensen, F. Polarization Consistent Basis Sets. III. The Importance of Diffuse Functions. *J. Chem. Phys.* **2002**, *117*, 9234-9240.
54. Frisch, M. J.; Trucks, G. W.; Schlegel, H. B.; Scuseria, G. E.; Robb, M. A.; Cheeseman, J. R.; Scalmani, G.; Barone, V.; Mennucci, B.; Petersson, G. A.; Nakatsuji, H.; Caricato, M.; Li, X.; Hratchian, H. P.; Izmaylov, A. F.; Bloino, J.; Zheng, G.; Sonnenberg, J. L.; Hada, M.; Ehara, M.; Toyota, K.; Fukuda, R.; Hasegawa, J.; Ishida, M.; Nakajima, T.; Honda, Y.; Kitao, O.; Nakai, H.; Vreven, T.; Montgomery Jr., J. A.; Peralta, J. E.; Ogliaro, F.; Bearpark, M. J.; Heyd, J.; Brothers, E. N.; Kudin, K. N.; Staroverov, V. N.; Kobayashi, R.; Normand, J.; Raghavachari, K.; Rendell, A. P.; Burant, J. C.; Iyengar, S. S.; Tomasi, J.; Cossi, M.; Rega, N.; Millam, N. J.; Klene, M.; Knox, J. E.; Cross, J. B.; Bakken, V.; Adamo, C.; Jaramillo, J.; Gomperts, R.; Stratmann, R. E.; Yazyev, O.; Austin, A. J.; Cammi, R.; Pomelli, C.; Ochterski, J. W.; Martin, R. L.; Morokuma, K.; Zakrzewski, V. G.; Voth, G. A.; Salvador, P.; Dannenberg, J. J.; Dapprich, S.; Daniels, A. D.; Farkas, Ö.; Foresman, J. B.; Ortiz, J. V.; Cioslowski, J.; Fox, D. J. *Gaussian 09*; Gaussian, Inc.: Wallingford, CT, **2009**.
55. Murray, C. W.; Handy, N. C.; Laming, G. J. Quadrature Schemes for Integrals of Density Functional Theory. *Mol. Phys.* **1993**, *78*, 997-1014.
56. Oviedo, M. B.; Ilawe, N. V.; Wong, B. M. Polarizabilities of π-Conjugated Chains Revisited: Improved Results from Broken-Symmetry Range-Separated DFT and New CCSD(T) Benchmarks. *J. Chem. Theory Comput.* **2016**, *12*, 3593-3602.
57. Lee, C.; Yang, W.; Parr, R. G. Development of the Colle-Salvetti Correlation-Energy Formula into a Functional of the Electron Density. *Phys. Rev. B* **1988**, *37*, 785-789.
58. Medvedev, M. G.; Bushmarinov, I. S.; Sun, J.; Perdew, J. P.; Lyssenko, K. A. Density Functional Theory is Straying from the Path Toward the Exact Functional. *Science* **2017**, *355*, 49-52.
59. Pari, S.; Wang, I. A.; Liu, H.; Wong, B. M. Sulfate Radical Oxidation of Aromatic Contaminants: A Detailed Assessment of Density Functional Theory and High-Level Quantum Chemical Methods. *Environ. Sci.: Processes Impacts*, **2017**, pubs.rsc.org ePrint archive. http://pubs.rsc.org/en/content/articlehtml/2017/em/c7em00009j (accessed Feb 26 2017).

60. Jensen, F. Polarization Consistent Basis Sets. III. The Importance of Diffuse Functions. *J. Chem. Phys.* **2002**, *117*, 9234-9240.
61. Kramida, A.; Ralchenko, Y.; Reader, J.; NIST ASD Team. NIST Atomic Spectra Database (ver. 5.2). National Institute of Standards Technology: Gaithersburg, MD, **2014**.
62. Rienstra-Kiracofe, J. C.; Tschumper, G. S.; Schaefer, H. F., 3rd; Nandi, S.; Ellison, G. B. Atomic and Molecular Electron Affinities: Photoelectron Experiments and Theoretical Computations. *Chem. Rev.* **2002**, *102*, 231-82.